\documentclass[a4paper,10pt,twoside]{cpc-hepnp}

\usepackage{multicol}
\usepackage{graphicx}
\usepackage{booktabs}
\usepackage{amssymb,bm,mathrsfs,bbm,amscd}
\usepackage[tbtags]{amsmath}
\usepackage{lastpage}
\usepackage{CJK}

\begin{document}

\fancyhead[c]{\small Chinese Physics C~~~Vol. 40, No. 5 (2016)
056202} \fancyfoot[C]{\small 010201-\thepage}

\footnotetext[0]{Received 17 9 2015}

\title{Neutron time-of-flight spectroscopy measurement \\using a waveform digitizer
\thanks{Supported by Strategic Priority Research Program of the Chinese Academy of Science(TMSR) (No.XDA02010100),
 National Natural Science Foundation of China(NSFC)(No.11475245,No.11305239),
Shanghai Key Laboratory of Particle Physics and Cosmology (11DZ2260700)} }

\author{%
      LIU Longxiang$^{1,2;1)}$\email{liulongxiang@sinap.ac.cn},%
\quad WANG Hongwei$^{1,2;2)}$\email{wanghongwei@sinap.ac.cn},%
\quad MA Yugang$^{1}$,
\quad CAO Xiguang$^{1,2}$,
\quad CAI Xiangzhou$^{1}$,\\
\quad CHEN Jingen$^{1}$,
\quad ZHANG Guilin$^{1}$,
\quad HAN Jianlong$^{1}$,
\quad ZHANG Guoqiang$^{1,2}$,
\quad HU Jifeng$^{1}$,\\
and \quad WANG Xiaohe$^{1}$
}
\maketitle

\address{%
$^1$ Shanghai Institute of Applied Physics, Chinese Academy of Sciences, Shanghai 201800, China\\
$^2$ Key Laboratory of Nuclear Radiation and Nuclear Technology, Chinese Academy of Science, Shanghai 201800, China\\
}

\begin{abstract}
The photoneutron source (PNS, phase 1), an electron linear accelerator (linac)-based pulsed neutron facility that uses the time-of-flight (TOF) technique, was constructed for the acquisition of nuclear data from the Thorium Molten Salt Reactor(TMSR) at the Shanghai Institute of Applied Physics (SINAP). The neutron detector signal
 used for TOF calculation, with information on the pulse arrival time, pulse shape, and pulse height, was recorded by using a waveform digitizer (WFD). By using the pulse height and pulse-shape discrimination (PSD) analysis to identify neutrons and $\gamma$-rays, the neutron TOF spectrum was obtained by employing a simple electronic design, and a new WFD-based DAQ system was developed and tested in this commissioning experiment. The DAQ system developed is characterized by a very high efficiency with respect to millisecond neutron TOF spectroscopy.
\end{abstract}




\footnotetext[0]{\hspace*{-3mm}\raisebox{0.3ex}{$\scriptstyle\copyright$}2013
Chinese Physical Society and the Institute of High Energy Physics
of the Chinese Academy of Sciences and the Institute
of Modern Physics of the Chinese Academy of Sciences and IOP Publishing Ltd}%

\begin{multicols}{2}

\section{Introduction}

Accelerator-based neutron sources are the most efficient for high-resolution measurements of microscopic neutron cross-sections,
for neutron energies ranging from thermal to several megaelectronvolts (MeV)  region,
and especially for the hundreds of electronvolts ($\sim$100 eV) resonance region.
Through the nuclear reactions of energetic electrons, protons or charged particles bombarding a heavy target can produce short bursts of neutrons with a broad continuous energy spectrum. Pulsed neutrons \cite{gelina,nelbe,pnf,pnf2,kurri}, based on an  electron linear accelerator (linac), are a powerful tool for effectively measuring energy-dependent cross-sections with high resolution by using the time-of-flight (TOF) technique, which covers neutron energies from the thermal to tens of MeVs range.
The electrons are stopped in a heavy target by the bremsstrahlung mechanism, and the radiation pulse in the detector is called the $\gamma$ flash. At the same time, neutrons are produced by the ($\gamma$,n) reaction with the target element. The neutron energy can be deduced from the formula $E_{n}[ev]=(72.3\times{L_{eff}}[m]/TOF[\mu{s}])^{2} $, where $L_{eff}$ is the effective flight path.

 This technique provides the basic information for studying neutron-nucleus interactions by measuring the neutron cross-sections.
 Precise measurements of neutron cross-sections are necessary for evaluating the neutron flux density
 and energy spectrum around a reactor and for the safety design of nuclear reactors.
The photoneutron source (PNS, phase 1), a nuclear data project, was initiated to construct infrastructure for nuclear data acquisition
for the thorium molten salt reactor (TMSR) at the Shanghai Institute of Applied Physics(SINAP) \cite{jiangmianheng}.
 The source consists of a 15 MeV electron linac, water-cooled tungsten (W) target (diameter = 60 mm and length = 48 mm)with a 10-cm-long polyethylene moderator, and a 5.7-m-long  effective TOF path.
 The nominal beam energy of the designed electron linac was 15 MeV with a maximal power of 7.5 kW, pulse frequency in the 1-266 Hz range, and  pulse widths of 3 ns, 30 ns, and 3 $\mu$s \cite{lin,lin2,wang,zhangmeng}.

For neutron TOF measurement, three issues must be addressed:
(1) handling long  neutron flight time (for example, the thermal neutron TOF value is 2.6 ms for a 5.7-m-long flight path,
much longer than the time for traditional TDC modules operating in the ns or $\mu$s range);
(2) detecting multiple neutron events generated by the source at the same time, different from the normal event-by-event recording mode;
(3) simplifying the electronics design.
The enhancement of the storage capability of hard disks and the rapid growth of computer performance has promoted
 the development of waveform digitizers (WFDs) and their application to nuclear experiments.
When a WFD is connected to a detector, all the information about the state of the detector's output can be collected during the experiment.
This new technique has obvious advantages compared with the traditional technique, in which
separate electronic modules are applied for selection and storage of the information.
Figs.~\ref{bd} (a) and (b) show the block diagram of the WFD-based data acquisition for the neutron TOF measurements at the PNS1
and the traditional measurement technique, respectively.
It is obvious that the WFD-based data acquisition reduces the complexity of the electronics compared with the traditional DAQ mode.
Some advantages of this method were presented in Refs.\cite{Kornilov,caen}.

  In this study, we demonstrate the application of this method by using a $^6$LiF(ZnS) scintillator \cite{EJ},
  coupled with a photomultiplier tube (PMT), for neutron TOF spectroscopy \cite{dulong,chang1,chang2}. Its efficiency for slow neutrons is about 70$\%$, calibrated using reactor thermal neutrons and a $^{252}Cf$ neutron source, and it is consistent with the Geant4 simulation result.
We demonstrate excellent timing and particle-type discrimination for the registered events, as well as good pulse amplitude resolution.

\begin{center}
\includegraphics[width=7cm]{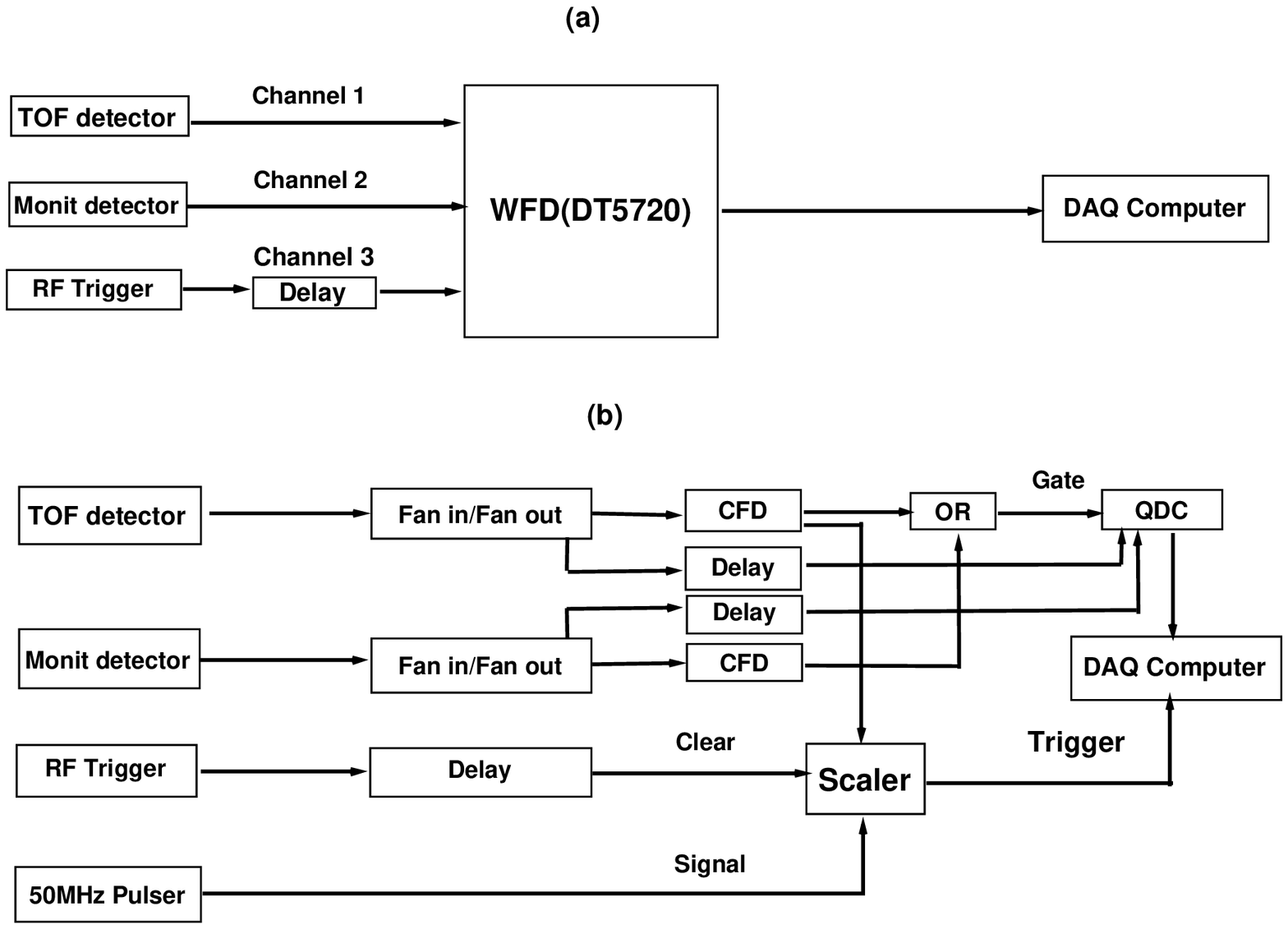}
\figcaption{\label{bd}    The block diagram of the data acquisition system for the neutron TOF measurements at the PNS1. (a) WFD-based data acquisition system.
 (b) Traditional data acquisition system. }
\end{center}

\section{ Experimental arrangement}

The experimental setup for the neutron TOF spectrum measurement is shown in Fig.~\ref{detector}.
The tungsten(W) target is located at a position where the electron beam hits the center of the target, and the target is
aligned vertically with the center of a TOF tube.
 A 5-cm-thick lead(Pb) plate with a diameter of 15 cm was placed at the entrance of the TOF tube to reduce the $\gamma$ flash
 generated by the electron burst in the target and scattered high-energy neutrons.
 Pb was chosen because of its low-energy cross-section,which is $\sim$11.48 barn at 0.007 eV and
11.38 barn above 0.2 eV \cite{pbcs}; therefore, the Pb block can serve as an effective low-band filter to allow easier filtering of faster neutrons than sub-thermal neutrons.
 The absorption samples of thermal neutrons were placed at the center of the TOF path.
 There were three different kinds of samples: 0.2-mm-thick  natural indium (In), 1-mm-thick cadmium (Cd), and a blank sample.

\begin{center}
\includegraphics[width=7cm]{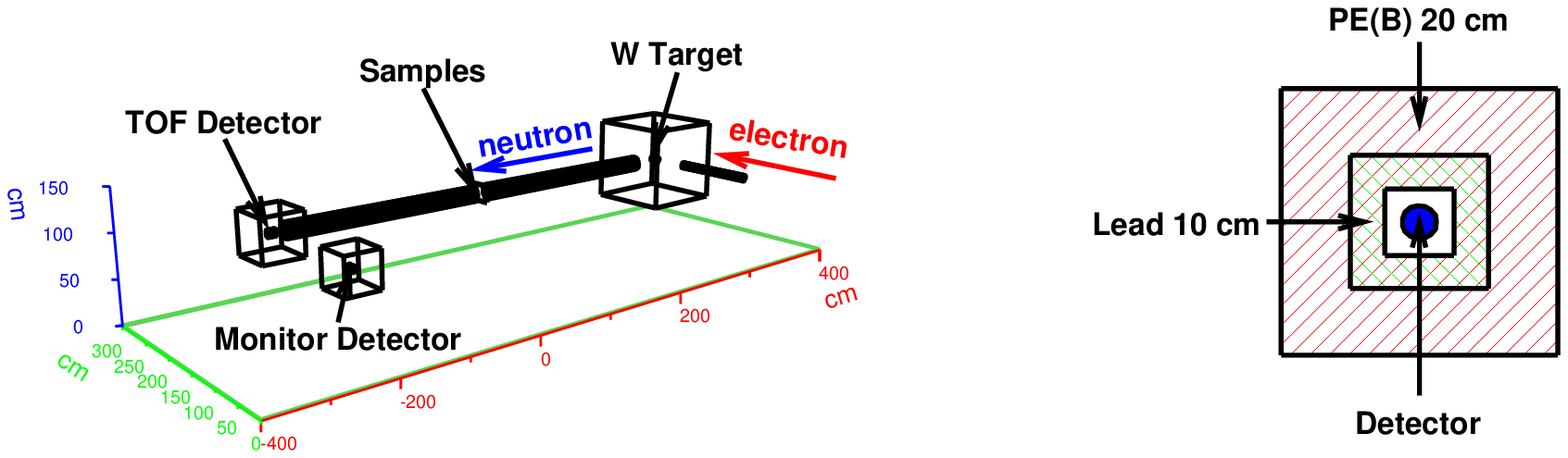}
\figcaption{\label{detector}   Experimental setup for the neutron TOF spectrum measurement(Left) and detector shielding cell(right). (color online) }
\end{center}

As a neutron detector, we used a $^6$LiF(ZnS) scintillator, product code EJ426HD2 from Eljen company, with a diameter of 50 mm and
thickness of 0.5 mm, mounted on an EMI9813 photomultiplier produced by ET enterprise Ltd.
There were two identical neutron detectors; one was located 5.7 m away from the photoneutron target for the neutron TOF spectrum measurement, and the other was at the bottom to the right of the neutron beam pipe for monitoring the beam intensity  during the experiment,as shown in Fig.~\ref{detector}(a).
Both the neutron detectors were shielded by 10 cm Pb bricks and 20 cm borated polyethylene plates(10$\%$ w.t.), as shown in Fig.~\ref{detector}(b).

\section{ Data acquisition}

The neutron energy spectra produced from the W(tungsten) target with a 10-cm-long polyethylene moderator were measured by using the TOF technique.
In this experiment, we used a CAEN DT5720 digitizer, four channel WFD with a bandwidth of 125 MHz,
 and 2 Vpp dynamic ranges on single-ended coax MCX input connectors.
It included an analog-to-digital converter (ADC) that sampled waveforms in a real-time mode at rates of 250 MS/s (mega
sample points per second) and converted them into a 12-bit-based code.
Its DC offset was calibrated by using a 16-bit DAC on each channel in the  $\pm 1$ V range.
The data stream was continuously stored in a circular memory buffer.
 When the trigger occurred, the FPGA stored an additional N samples for the post-trigger phase and locked the buffer, which could then be read using a USB or optical link.
 The acquisition continued into a new buffer, without dead time \cite{caen}.
In the present case, waveform traces with 512 samples  were stored for off-line analysis,
and the post-trigger phase waveform comprised 80$\%$ of the entire acquisition window.

The signal obtained by the neutron detector for TOF measurement was connected to channel 1 of the DT5720, and the signal for
 monitoring the beam intensity was connected to channel 2.
The trigger signal (the electron GUN start signal) was connected to channel 3, as the start signal for TOF calculation.
 The DT5720 was connected to a personal computer with a Linux operating system via an optical cable.
 The data were collected, stored, and analyzed on this computer.
 Because of different baselines, the thresholds of the three channels were set as -14.65 mV, -4.88 mV ,
 and -4.88 mV respectively. The self-trigger mode was used, which means that any channel signal going above the threshold will trigger the DAQ.
 The trigger occurs on the falling edge of the signal.
The data acquisition software we developed based on ROOT \cite{root} and Gnuplot \cite{gnuplot} is known as TMSR-Digitizer-DAQ.
DT5720 houses USB 2.0 and optical link interfaces. USB 2.0 allows data transfers up to 30 MBytes/s. The optical link supports a transfer rate of 80 MBytes/s, and offer daisy-chain capability. Therefore it is possible to connect up to 8 ADC modules(total 32 channels) to an A2818 Optical Link PCIe Controller or 32 modules(total 128 channels) to an A3818(for VME digitizer version,the maximum is 32$\times$16=512 channels).
It is a very high efficiency DAQ system for millisecond or second level neutron TOF spectroscopy, which incorporates data compression,
automatic change of target, and  automatic file opening functions.

During the commissioning experiment, the electron linac was operated with a repetition rate of 40 Hz, a pulse
width of 1.5 $\mu$s, a peak current of 10 $\mu$A, and an electron energy of 15 MeV.

\section{Data analysis}

 The TOF spectrum was only measured in the direction normal to the incident electron beam.
 The method applied for n/$\gamma$ identification for TOF spectrum calculations was the pulse-shape discrimination
 (PSD, calculated by using integral lengths) method that was used for identifying neutron and $\gamma$-rays \cite{chang1}.
 There are two integral lengths for every waveform trace: one is the short gate $Qs$ ($\sim$50ns),
 and the other one is the long gate $Ql$ ($\sim$240ns).
 The integral lengths of $Qs$ and $Ql$ depend on the pulse width of the $\gamma$-rays and neutrons.
 The two-dimensional distribution of $Ql$ versus PSD=$(Ql-Qs)/Ql$ is shown in Fig.~\ref{psd}.
 Neutrons are selected for the following conditions: $PSD>0.4$ and $Ql>350$.
\begin{center}
\includegraphics[width=8cm]{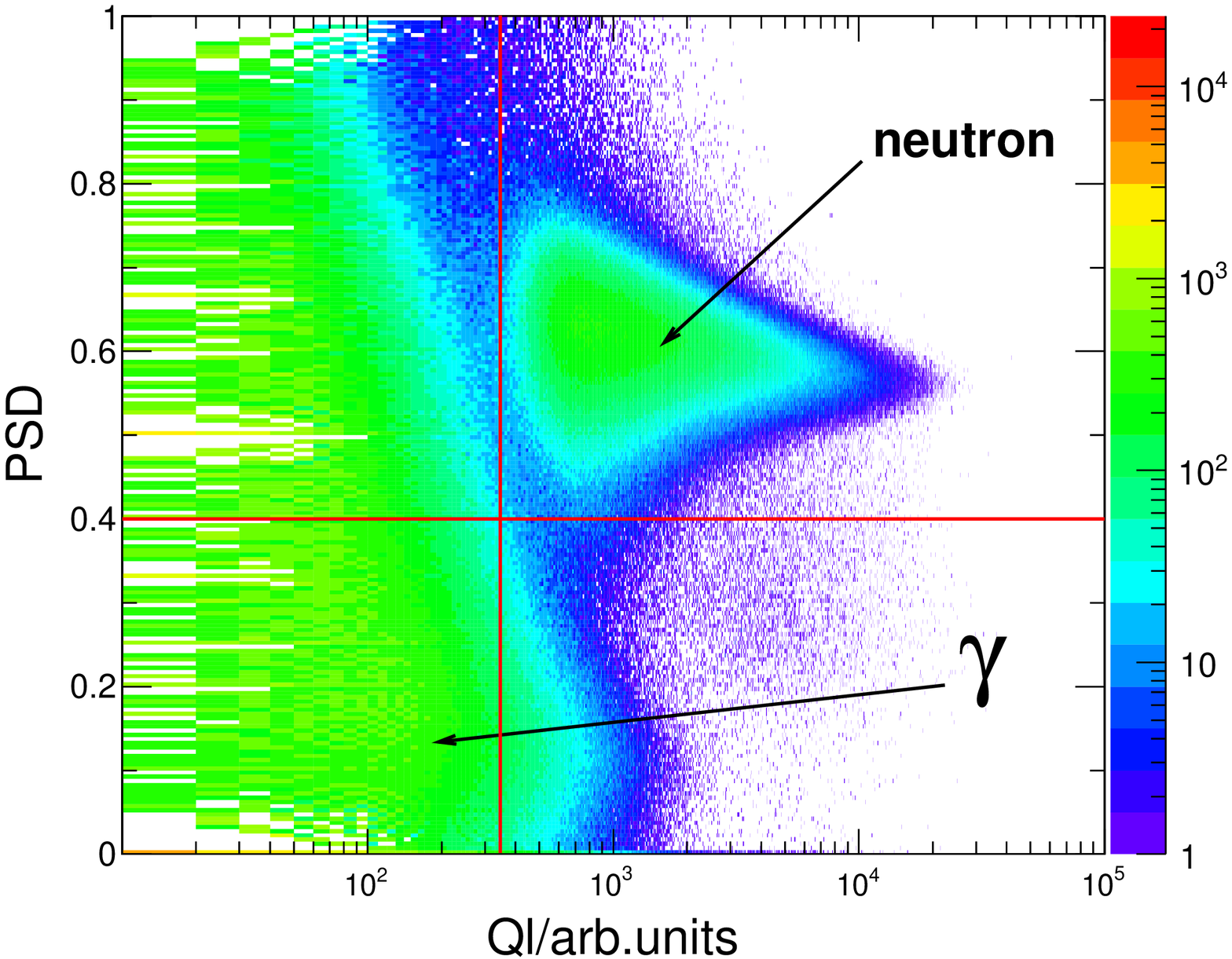}
\figcaption{\label{psd}   Two-dimensional distribution of detector events.(color online) }
\end{center}

The $\gamma$ flash is generated by the electron burst in the target.
The $\gamma$ flash and the scattered neutron waveform of the TOF detector were captured online and
are shown in Fig.~\ref{ch3} at the recording time of 32 $\mu$s.
The $\gamma$ flash with a long tail leads to a dead time of $\sim$10 $\mu$s and can be recognised as a very big pulse, then with some small neutron and $\gamma$ pulses, The dead time corresponds to the neutron energy of $\sim$1.6 keV
for a 5.7-m-long effective flight path, which reduces the measurable range of neutron energies, so the effective neutron energy is in the thermal to 1 keV range.

\begin{center}
\includegraphics[width=8cm]{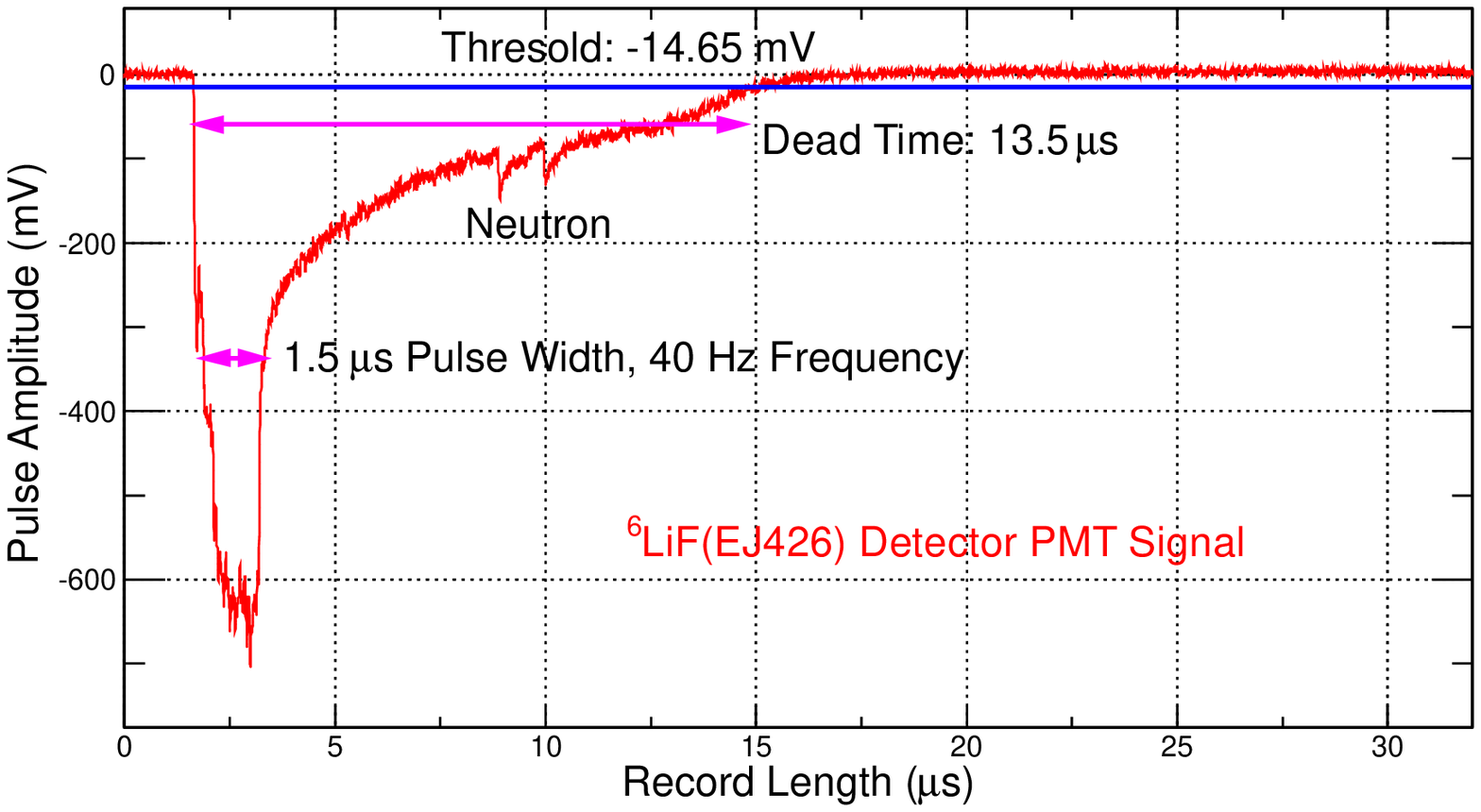}
\figcaption{\label{ch3} The $\gamma$ flash waveform of the TOF detector with the recording time of 32 $\mu$s. (color online)}
\end{center}


Fig.~\ref{wave6} shows the waveforms for six different recordings of the three channels after the arrival of one $\gamma$ flash.
Here, the normal record time was 2048 ns for neutron multi-event recordings.
TTT is the trigger time tag, which is a 31-bit counter (31-bit count + 1-bit as a rollover flag) and is
incremented every two ADC clock cycles. It represents the reference time of the trigger.
The TTT resolution is 16 ns and it can capture up to 17 s (i.e., $8 ns\times(2^{31}-1)$).
If any of the three channels are triggered, all of them will record one event.
Therefore, their TTTs will be identical.
As shown in Fig.~\ref{wave6}, the first column is triggered by channel 3 and the second, third, and fourth
columns are triggered by channel 2. All the columns are triggered by the $\gamma$ flash with a long tail,
and they are not neutron or $\gamma$ signals.
The fifth column is triggered by channel 1, and it is a neutron signal;
 the sixth column is triggered by channel 2, and is also a neutron signal.

\end{multicols}
\ruleup
\begin{center}
\includegraphics[width=16.5cm]{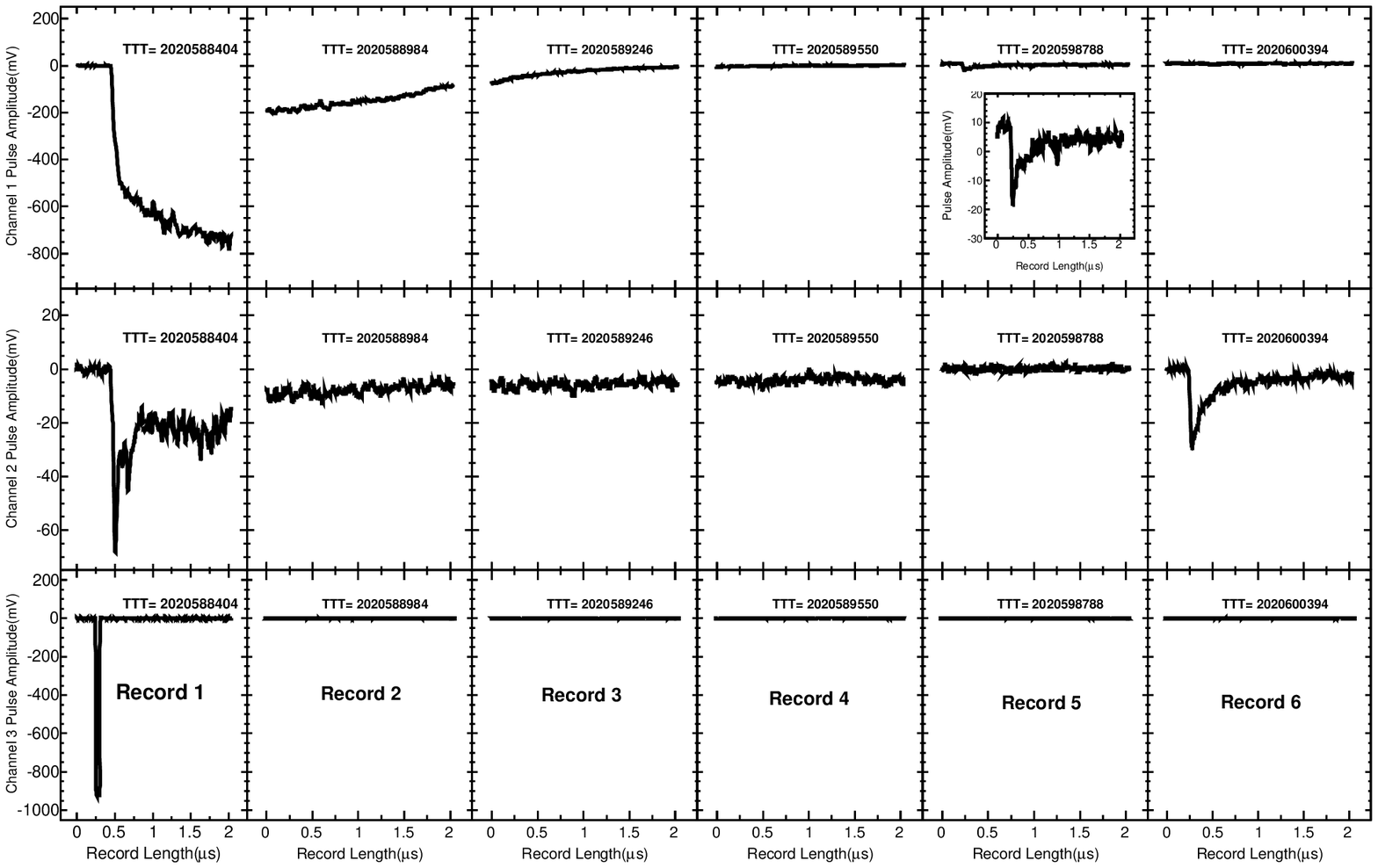}
\figcaption{\label{wave6} Six waveform records for three channels after one $\gamma$ flash (rows denote the channels; TTT is the trigger time tag).}
\end{center}
\ruledown

\begin{multicols}{2}

\section{TOF spectrum}

The falling edge of the channel 3 signal was used as the start signal and the neutron peak position of channel 1
was used as the stop signal for neutron TOF measurement.
Fig.~\ref{tof} shows the  neutron TOF spectra for the sample-in and open beam; (a) is the digitizer DAQ results and (b) is the traditional DAQ result in different accelerator operating modes. The two methods have the nearly same TOF shape.

They yield resonance peaks for In and Cd at 1.457 eV and 0.17 eV,
and the TOF values are 341$\pm$10 $\mu$s and 980$\pm$10  $\mu$s respectively for a 5.7-m-long neutron effective flight path.
 These results also show a large neutron background contribution from the environmental scattered neutrons, which come from the  radiation when electron beam bending is 90$^{\circ}$, and partly leak from the electron pipeline entrance.
 It will be reduced in the future by using shielding.
For each sample, the data were collected for 40 min.

Fig.~\ref{InCd} shows the neutron absorption spectra for the open beam subtracted from those of the  natural In ($^{113}$In, $^{115}$In) and Cd samples. By using the WFD-based TOF 
\begin{center}
\includegraphics[width=7cm]{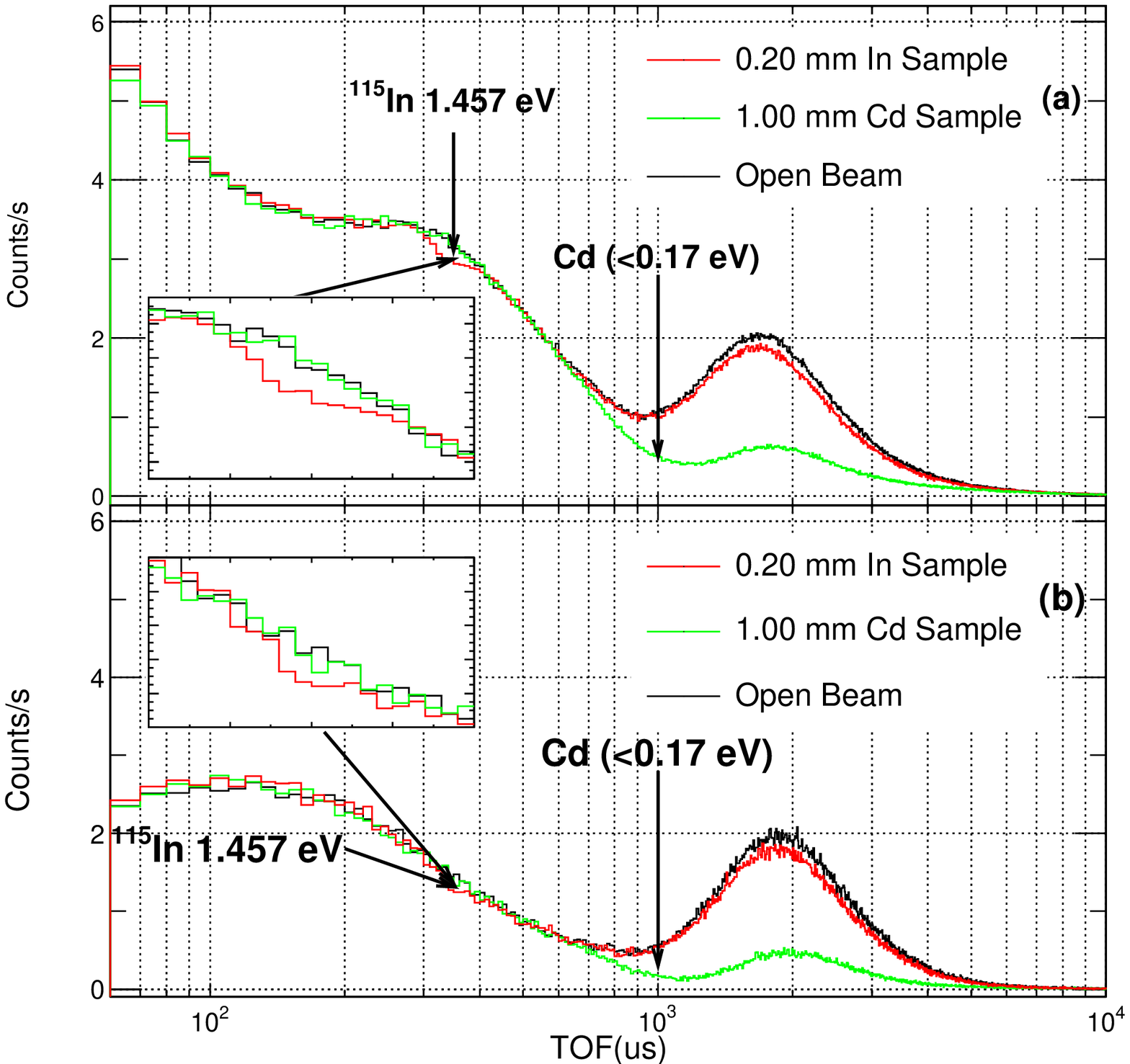}
\figcaption{\label{tof}    Neutron TOF spectra for the sample-in and open beam,(a) Digitizer method, (b) Traditional method. (color online)}.
\end{center}
\begin{center}
\includegraphics[width=7cm]{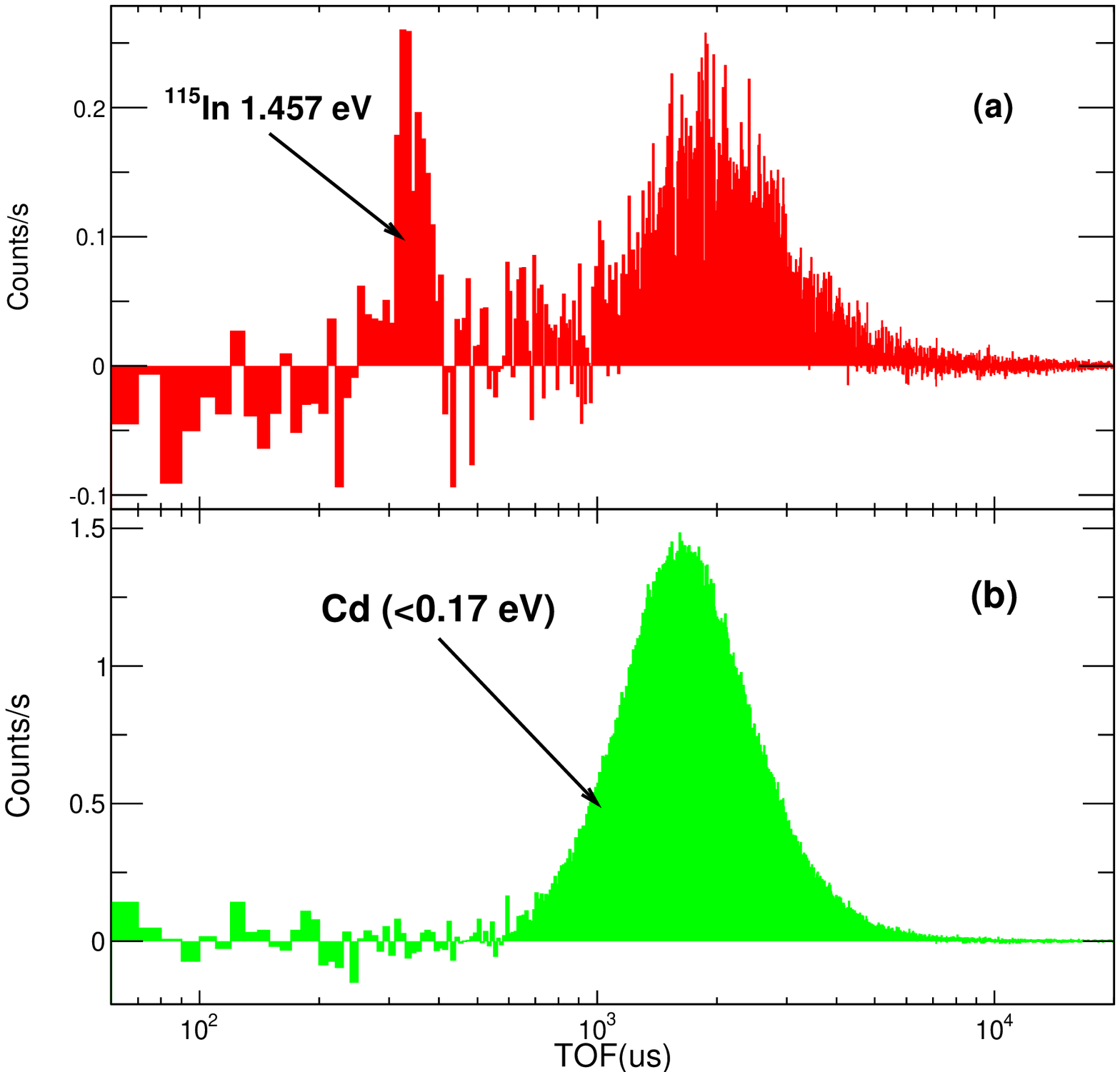}
\figcaption{\label{InCd}  Neutron absorption spectra of the In(a) and Cd(b) samples subtracted from the open beam. (color online)}
\end{center}
method, we obtained the resonance energies for In and Cd as 1.457 eV and 0.17 eV, respectively.
Clear peaks are observed in the In and Cd spectra.

\section{Conclusion}

In conclusion, we have demonstrated the application of a WFD  for neutron TOF spectroscopy.
 We investigated the PSD between neutrons and $\gamma$-rays, and performed the waveform data analysis method.
 Based on the WFD and DAQ system, the millisecond TOF spectra deduced from the time tag of $\gamma$ flash and neutron pulse
 are in good agreement with the results obtained by using pulser+scaler methods.
In addition, the method allowed us to solve the millisecond TOF spectra and multi-event problems that occur in neutron experiments.
By using this method, we measured the resonance peaks for In at 1.457 eV and Cd cut-off from 0.17 eV to thermal neutron,
and then deduced that the TOF values were 341 $\mu$s and 980 $\mu$s respectively for a 5.7-m-long neutron effective flight path,
which are consistent with the theoretically calculated results.
In this measurement method, two problems were identified:
one is the $\gamma$ flash with a long tail, and the other one is a high neutron background level.
In the future, we will take steps to reduce the $\gamma$ flash tail and the neutron background.
This will help in achieving precise measurements of neutron cross-sections.

\emph{We are grateful to the linac accelerator group and to other faculty members, and appreciate their encouragement.}

\end{multicols}
\vspace{-1mm}
\centerline{\rule{80mm}{0.1pt}}
\vspace{2mm}

\begin{multicols}{2}

\end{multicols}

\clearpage


\begin{thebibliography}{90}

\vspace{3mm}

\bibitem{gelina} GELINA https://ec.europa.eu/jrc/en/research-facility/linear-electron-accelerator-facility

\bibitem{nelbe} nELBE http://www.hzdr.de/db/Cms?pNId=145

\bibitem{pnf} Kim G N, Lee Y S, Skoy V, et al. First experiment at the Pohang neutron facility. Journal of the Korean Physical Society, 2001, 38: 14---18

\bibitem{pnf2} Kim G N, Kovalchuk V, Lee Y S, et al. Measurement of photoneutron spectrum at Pohang neutron facility.Nuclear Instrument and Methods A, 2002, 485: 458---467

\bibitem{kurri} KURRI http://www.rri.kyoto-u.ac.jp/en/facilities/ela, 2014

\bibitem{jiangmianheng} JIANG Mianheng, XU Hongjie, DAI Zhimin. Advanced fission energy program-TMSR nuclear energy system.Bulletin of the Chinese Academy of Sciences, 2012, 27(3):366---374

\bibitem{lin} Lin Z K, Sun G M, Chen J G, et al. Simulation and optimization for a 30-MeV electron accelerator driven neutron source. Nuclear Science and Techniques, 2012,23: 272---276

\bibitem{lin2} Lin Z K, Zou X, Cao Y, et al. Analysis of simulation for neutron target driven by 15-MeV electron beam.Atomic Energy Science and Technology, 2012, 46(Suppl):26---30

\bibitem{wang} WANG Hongwei, CAI Xiangzhou, CHEN Jingen et al. NUCLEAR TECHNIQUES, 2014, {\bf 37}(10): 1---5

\bibitem{zhangmeng} Zhang M, Li X, Fang W C, et al. LINAC design for nuclear data measurement facility. Proceedings of IPAC2013, Shanghai, China

\bibitem{Kornilov} N.V. Kornilov et al. Nuclear Instruments and Methods in Physics Research A 497 (2003) 467---478

\bibitem{caen} CAEN http://www.caen.it/

\bibitem{EJ} Eljen http://www.eljentechnology.com/

\bibitem{dulong} DU Long, CHANG Le, WANG Yuting, et al. Detection efficiency simulation and measurement of 6LiI/natLiI scintillation detector(in Chinese). Nuclear Techniques, 2014, 37(4):040201

\bibitem{chang1} CHANG Le, LIU Yingdu, DU Long, et al. Pulse shape discrimination and energy calibration of EJ301 liquid scintillation detector(in chinese). Nuclear Techniques, 2015, 38(2):020501

\bibitem{chang2} CHANG Le, LIU Longxiang, WANG Hongwei, et al.Double pulse waveform spectrum of EJ339A capture-gated neutron detector(in chinese). Nuclear Techniques, 2015, 38(5):050403

\bibitem{pbcs} http://t2.lanl.gov/nis/data/endf/endfvi-n.html

\bibitem{root} ROOT http://root.cern.ch/drupal/

\bibitem{gnuplot} GNUPLOT http://www.gnuplot.vt.edu/

\end{thebibliography}
\end{document}